\documentclass[twocolumn,aps,prd,superscriptaddress,nofootinbib,preprintnumbers,showpacs,floatfix]{revtex4-2}
\usepackage{amsmath,amssymb,amsfonts}
\usepackage{graphicx}
\usepackage{hyperref}
\usepackage{xcolor}
\usepackage{bm}

\begin{document}
	
\title{Thermodynamics of thin‑shell wormholes}

\author{Francisco S. N. Lobo} \email{fslobo@ciencias.ulisboa.pt}
\affiliation{Instituto de Astrof\'{i}sica e Ci\^{e}ncias do Espa\c{c}o, Faculdade de Ci\^{e}ncias da Universidade de Lisboa, Edifício C8, Campo Grande, P-1749-016 Lisbon, Portugal}
\affiliation{Departamento de F\'{i}sica, Faculdade de Ci\^{e}ncias da Universidade de Lisboa, Edif\'{i}cio C8, Campo Grande, P-1749-016 Lisbon, Portugal}
\author{Manuel E. Rodrigues} \email{esialg@gmail.com}
\affiliation{Faculdade de F\'{i}sica, Programa de P\'{o}s-Gradua\c{c}\~{a}o em F\'{i}sica, Universidade Federal do Par\'{a}, 66075-110, Bel\'{e}m, Par\'{a}, Brazill}
\affiliation{Faculdade de Ci\^{e}ncias Exatas e Tecnologia, Universidade Federal do Par\'{a}, Campus Universit\'{a}rio de Abaetetuba, 68440-000, Abaetetuba, Par\'{a}, Brazil}
\date{\LaTeX-ed \today}

\begin{abstract}

We develop a unified thermodynamic description of dynamic thin-shell wormholes, starting from the transparent (vacuum bulk) case and then relaxing the transparency condition to include bulk matter crossing the throat. For isolated shells, we show that the entropy is conserved under transparent evolution, $TdS=0$, which holds for arbitrary dynamical motion of the throat within the assumptions of the thin-shell formalism. When bulk matter is present, the generalised first law becomes $T\dot{S}=A\Phi$, where $\Phi$ is the net energy flux; entropy increases (decreases) when matter flows into (out of) the shell.
Explicit expressions for the flux are provided for null dust and massless scalar fields, and quantum pair production (Hawking-like emission) is also discussed. A formulation of the generalised second law is presented, and it is consistent with standard thermodynamic expectations under conditions where heat flows from hotter to colder regions. As a concrete astrophysical application, we study accretion of null dust: the critical accretion rate above which the throat becomes dynamically unstable depends sensitively on the spacetime geometry and the shell equation of state, highlighting the environmental dependence of these configurations.
Fluctuations in the energy flux induce a stochastic force on the throat dynamics, leading to a Langevin description with an associated fluctuation-dissipation relation. This results in diffusion of the throat radius and additional entropy production. The framework applies to a broad class of spherically symmetric thin-shell constructions within general relativity. Our results provide evidence that thin-shell wormholes admit a consistent thermodynamic description, quantify their sensitivity to accretion processes, and suggest possible avenues for probing exotic compact objects with future gravitational wave observations.

\end{abstract}

	\maketitle



\section{Introduction}

The discovery that black holes possess entropy and temperature, culminating in the laws of black hole mechanics~\cite{Bardeen:1973gs} and the quantum emission of Hawking radiation~\cite{Hawking:1975vcx}, established a profound link between gravity, quantum field theory, and thermodynamics. The insight that a black hole's entropy is proportional to its horizon area~\cite{Bekenstein:1973ur} opened the door to a microscopic interpretation of gravitational degrees of freedom. Since then, the thermodynamic properties of horizons have become a cornerstone of theoretical physics, influencing developments ranging from the holographic principle~\cite{Ryu:2006bv,Maldacena:2013xja} to the quantum nature of spacetime.

Traversable wormholes, introduced in the seminal works~\cite{Morris:1988cz,Morris:1988tu}, represent another class of compact objects that, while sharing many geometric features with black holes, lack an event horizon. Their traversability imposes stringent conditions on the matter that supports the throat, in particular the violation of the null energy condition (NEC) \cite{Visser:1995cc,Lobo:2017cay}. Because traversable wormholes require NEC violation within general relativity, their existence typically calls for either quantum effects \cite{Ford:1995wg,Sushkov:2009hk} or modifications of gravity theories \cite{Harko:2013yb}. A particularly elegant and systematic method to construct such geometries is the thin-shell (or cut-and-paste) formalism, developed in~\cite{Visser:1989kh,Visser:1989kg}. In this approach, two copies of a given spacetime are excised at a timelike hypersurface and identified, producing a geodesically complete manifold whose junction surface $\Sigma$ becomes the wormhole throat. The matching conditions for singular hypersurfaces, developed in the works~\cite{Israel:1966rt,Lanczos}, determine the surface stress-energy tensor of the shell from the jump in extrinsic curvature. Linearised stability analyses of such thin-shell wormholes have been performed for a variety of backgrounds~\cite{Poisson:1995sv,Eiroa:2003wp,Lobo:2003xd,Ishak:2001az}.

Remarkably, the thermodynamic properties of thin shells have received comparatively less attention. Early studies~\cite{York:1986it,Hiscock:1989uj}, building on the Noether charge approach~\cite{Wald:1993nt}, derived the first law for static thin shells and established the constancy of entropy along equilibrium sequences. However, a fully systematic treatment of genuinely dynamical configurations that includes throat motion, time-dependent temperature, non-equilibrium processes (such as evaporation or accretion), and relaxation of the transparency condition is still incomplete. Moreover, the Bekenstein bound~\cite{Bekenstein:1980jp} provides a universal constraint on entropy that should also constrain dynamic throats, while quantum corrections to the entropy (such as logarithmic terms~\cite{Sen:2012dw}) are expected to modify classical adiabatic behaviour on long timescales. More recently, regularised black hole metrics that are free of curvature singularities, such as the T-duality inspired solution~\cite{Nicolini:2019irw}, the Hayward metric~\cite{Hayward:2005gi}, and the Simpson-Visser black bounce~\cite{Simpson:2018tsi}, have been proposed as useful backgrounds for wormhole constructions. The generic stability analysis of dynamic thin shells in standard general relativity was unified in~\cite{Garcia:2011aa} and extended to gravastars in~\cite{MartinMoruno:2011rm}. The Gibbs-Duhem relation for relativistic fluids, as discussed in~\cite{Quevedo:1994yg}, provides the thermodynamic identities required when bulk matter crosses the throat.

In the present work we develop a comprehensive thermodynamic framework for dynamic thin-shell wormholes. Starting from the transparent (vacuum bulk) case, we show that the shell entropy behaves as an adiabatic invariant along dynamical trajectories under the absence of flux across the shell, thereby generalising equilibrium results to time-dependent motion of the throat. We define an effective dynamical temperature in terms of a suitably generalised notion of surface gravity, verify the Bekenstein bound in this setting, analyse non-equilibrium evaporation, and compute leading quantum corrections. We then relax the transparency condition, allowing bulk matter to cross the throat. This introduces a source term $\Phi = [T_{\mu\nu}U^\mu n^\nu]^{+}_{-}$ in the surface Bianchi identity, leading to a generalised first law $T\dot{S}=A\Phi$. We provide explicit expressions for $\Phi$ for null dust, massless scalar fields, and quantum pair production (Hawking-like emission). We formulate a generalised second law, analyse thermalisation with a heat bath, and study the accretion of null dust, including an estimate of the critical accretion rate for dynamical instability. Finally, we incorporate fluctuations of the flux, deriving a Langevin equation and a fluctuation-dissipation relation, which leads to stochastic diffusion of the throat and additional entropy production.

The paper is organised as follows. Section~\ref{sec:thinshell} reviews the thin-shell formalism and defines the dynamical surface stresses. Section~\ref{sec:thermodynamics_dynamic} establishes the first law and adiabatic invariance for transparent wormholes, introduces the effective dynamical temperature, verifies the Bekenstein bound, and discusses evaporation and quantum corrections. Section~\ref{sec:non_transparent} relaxes the transparency condition, deriving the modified conservation equation, the generalised first law, and explicit flux expressions. In Sec.~\ref{sec:applications}, we formulate the generalised second law, analyse thermalisation, and study an accretion example involving null dust; stochastic effects are also investigated. Our conclusions and outlook are presented in Section~\ref{sec:conclusion}.

\section{Thin-shell formalism}
\label{sec:thinshell}

We consider two distinct, static, spherically symmetric spacetimes
\(\mathcal{M}_\pm\) with metrics
\begin{equation}
	ds_\pm^2 = -f_{\pm}(r_\pm) dt_\pm^2 + f_{\pm}^{-1}(r_\pm) dr_\pm^2
	+ r_\pm^2 d\Omega^2,
	\label{eq:bulk_metric}
\end{equation}
where \(d\Omega^2 = d\theta^2 + \sin^2\theta\, d\phi^2\) is the metric
on the unit two-sphere, and \(f_{\pm}(r_\pm)\) are arbitrary metric
functions. The wormhole throat \(\Sigma\) is constructed by
excising the inner regions \(r_\pm < a(\tau)\) and identifying the two
timelike boundaries \(r_\pm = a(\tau)\). The resulting manifold
\(\mathcal{M} = \mathcal{M}_+ \cup \mathcal{M}_-\) is geodesically
complete provided that the excision removes any singular regions or
horizons. The throat is parametrised by coordinates \(\xi^i = (\tau, \theta, \phi)\),
where \(\tau\) is the proper time of a comoving observer. The induced
metric on \(\Sigma\) is
\begin{equation}
	ds_\Sigma^2 = -d\tau^2 + a(\tau)^2 d\Omega^2 .
	\label{eq:induced_metric}
\end{equation}


The embedding of \(\Sigma\) into \(\mathcal{M}_\pm\) is given by
\(x^\mu_\pm = (t_\pm(\tau), a(\tau), \theta, \phi)\). The condition
that \(\tau\) is proper time on \(\Sigma\) determines \(\dot{t}_\pm\):
\begin{equation}
	\dot{t}_\pm = \frac{\sqrt{f_{\pm}(a)+\dot{a}^2}}{f_{\pm}(a)},
\end{equation}
where the overdot denotes \(d/d\tau\). The unit normal one-form
to \(\Sigma\) (pointing from \(\mathcal{M}_-\) to \(\mathcal{M}_+\))
is
\begin{equation}
	n_\mu^\pm = \pm \left( -\dot{a},\; \frac{\sqrt{f_{\pm}(a)+\dot{a}^2}}{f_{\pm}(a)},\; 0,\; 0 \right),
	\label{eq:normal}
\end{equation}
with the upper sign for the \(+\) side and the lower sign for the
\(-\) side. One verifies orthogonality \(n_\mu^\pm \partial x^\mu_\pm/\partial \xi^i = 0\) and normalization
\(n_\mu^\pm n^{\mu\,\pm}=+1\) (the normal is spacelike).

\subsection{Extrinsic curvature}

The extrinsic curvature of \(\Sigma\) as embedded in \(\mathcal{M}_\pm\) is
\begin{equation}
	K_{ij}^\pm = -n_\mu^\pm \left( \frac{\partial^2 x^\mu_\pm}{\partial \xi^i \partial \xi^j}
	+ \Gamma^{\mu}_{\alpha\beta} \frac{\partial x^\alpha_\pm}{\partial \xi^i}
	\frac{\partial x^\beta_\pm}{\partial \xi^j} \right),
	\label{eq:extrinsic_definition}
\end{equation}
where \(\Gamma^{\mu}_{\alpha\beta}\) are the Christoffel symbols of
\(g_{\mu\nu}^\pm\). Because of spherical symmetry, \(K_{ij}^\pm\) is diagonal,
with non-vanishing angular and temporal components.


For \(\theta\) and \(\phi\), the embedding functions are simply
\(\theta = \xi^2\), \(\phi = \xi^3\). The second derivatives vanish,
so only the Christoffel term contributes. Using
\(\Gamma^r_{\theta\theta} = -r f_{\pm}(r)\) for the metric (1), we find
\begin{equation}
	K_{\theta\theta}^\pm = -n_r^\pm \bigl(-a f_{\pm}(a)\bigr) = a f_{\pm}(a) n_r^\pm .
\end{equation}
With \(n_r^\pm = \pm \frac{\sqrt{f_{\pm}+\dot{a}^2}}{f_{\pm}}\), we obtain
\(K_{\theta\theta}^\pm = \pm a \sqrt{f_{\pm}(a)+\dot{a}^2}\). Raising the
index with \(h^{\theta\theta}=1/a^2\) gives
\begin{equation}
	K^\theta_{\;\theta\,\pm} = \pm \frac{1}{a} \sqrt{f_{\pm}(a)+\dot{a}^2}.
	\label{eq:Ktheta_general}
\end{equation}


A direct calculation (or standard result for spherically symmetric
thin shells) yields
\begin{equation}
	K^\tau_{\;\tau\,\pm} = \pm \frac{\ddot{a} + \frac{1}{2} f'_{\pm}(a)}{\sqrt{f_{\pm}(a)+\dot{a}^2}} .
	\label{eq:Ktau_general}
\end{equation}
The derivation uses the explicit form of \(\dot{t}_\pm\) and the
Christoffel symbols.

\subsection{The Lanczos equations}

The jump of the extrinsic curvature across \(\Sigma\) is defined as
\([K_{ij}] = K_{ij}^+ - K_{ij}^-\). Using Eqs.~(\ref{eq:Ktheta_general}) and (\ref{eq:Ktau_general}) we obtain
\begin{align}
	[K^{\theta}_{\;\theta}] &= K^{\theta}_{\;\theta,\,+} - K^{\theta}_{\;\theta,\,-}
	\nonumber \\
	&= \frac{1}{a}\left( \sqrt{f_{+}(a)+\dot{a}^2} + \sqrt{f_{-}(a)+\dot{a}^2} \right),
	\label{eq:jump_theta_general}
	\\[4pt]
	[K^{\tau}_{\;\tau}] &= K^{\tau}_{\;\tau,\,+} - K^{\tau}_{\;\tau,\,-}
	\nonumber \\
	&= \frac{\ddot{a} + \frac{1}{2} f'_{+}(a)}{\sqrt{f_{+}(a)+\dot{a}^2}}
	+ \frac{\ddot{a} + \frac{1}{2} f'_{-}(a)}{\sqrt{f_{-}(a)+\dot{a}^2}} .
	\label{eq:jump_tau_general}
\end{align}
The trace of the jump is \([K] = h^{ij}[K_{ij}] = [K^\tau_{\;\tau}] + 2[K^\theta_{\;\theta}]\).

The Israel-Lanczos junction conditions relate the jump to the
surface stress-energy tensor \(S_{ij}\):
\begin{equation}
	S_{ij} = -\frac{1}{8\pi} \left( [K_{ij}] - h_{ij} [K] \right).
	\label{eq:Lanczos}
\end{equation}

For a spherically symmetric shell, \(S^{i}_{\;j} = \text{diag}(-\sigma, \mathcal{P}, \mathcal{P})\), where \(\sigma\) is the surface energy density and \(\mathcal{P}\) the tangential pressure. Substitution yields
\begin{align}
	\sigma &= -\frac{1}{4\pi}\,[K^\theta_{\;\theta}], \\
	\mathcal{P} &= \frac{1}{8\pi}\left( [K^\tau_{\;\tau}] + [K^\theta_{\;\theta}] \right).
\end{align}
Inserting the explicit jumps gives
\begin{equation}
	\sigma(a,\dot{a}) = -\frac{1}{4\pi a}\left( \sqrt{f_{+}(a)+\dot{a}^2} + \sqrt{f_{-}(a)+\dot{a}^2} \right),
	\label{eq:sigma_general}
\end{equation}
\begin{eqnarray}
	\mathcal{P}(a,\dot{a},\ddot{a}) &=&
	\frac{1}{8\pi} \Bigg[
	\frac{\ddot{a} + \frac{1}{2} f'_{+}(a)}{\sqrt{f_{+}(a)+\dot{a}^2}}
	+ \frac{\ddot{a} + \frac{1}{2} f'_{-}(a)}{\sqrt{f_{-}(a)+\dot{a}^2}}
	\nonumber \\
	&&+ \frac{1}{a}\left( \sqrt{f_{+}(a)+\dot{a}^2} + \sqrt{f_{-}(a)+\dot{a}^2} \right)
	\Bigg].
	\label{eq:P_general}
\end{eqnarray}
These expressions reduce to the symmetric case when \(f_{+}=f_{-}=f\).

\subsection{Conservation equation for the shell}

When the bulk spacetimes are vacuum (\(T_{\mu\nu}^{\text{bulk}}=0\))
and no energy flux crosses the shell, the contracted Gauss-Codazzi
identities imply conservation of the surface stress-energy tensor:
\begin{equation}
	S^{i}_{\;j|i} = 0,
\end{equation}
where the vertical bar denotes covariant differentiation with respect
to the induced metric \(h_{ij}\). For a perfect-fluid shell on the
induced metric (\ref{eq:induced_metric}), this conservation law reduces
to the single equation:
\begin{equation}
	\dot{\sigma} + \frac{2\dot{a}}{a}\bigl(\sigma+\mathcal{P}\bigr) = 0 .
	\label{eq:conservation_dynamic}
\end{equation}
This relation holds for both symmetric and asymmetric configurations,
and follows purely from the intrinsic geometry of the shell in the
absence of bulk matter flux across the throat. It is independent of
any assumed equation of state, which may be imposed separately to
close the system. One can show by direct substitution that the explicit
expressions in Eqs.~(\ref{eq:sigma_general})--(\ref{eq:P_general})
satisfy Eq.~(\ref{eq:conservation_dynamic}) identically.

\subsection{Static limit and symmetric specialisation}

In the static limit \(\dot{a}=\ddot{a}=0\), the general expressions
reduce to
\begin{align}
	\sigma_0 =& -\frac{1}{4\pi a_0}\left( \sqrt{f_{+}(a_0)} + \sqrt{f_{-}(a_0)} \right), \\
	\mathcal{P}_0 =& \frac{1}{8\pi} \Bigg( \frac{f'_{+}(a_0)}{2\sqrt{f_{+}(a_0)}}
	+ \frac{f'_{-}(a_0)}{2\sqrt{f_{-}(a_0)}}
	\nonumber \\
	&+ \frac{1}{a_0}\bigl( \sqrt{f_{+}(a_0)} + \sqrt{f_{-}(a_0)} \bigr) \Bigg).
\end{align}
For a symmetric wormhole (\(f_{+}=f_{-}=f\)), these reduce to the known
static results \(\sigma_0 = -\frac{1}{2\pi a_0}\sqrt{f(a_0)}\) and
\(\mathcal{P}_0 = \frac{1}{4\pi a_0}\,
\frac{f(a_0)+\frac{a_0}{2}f'(a_0)}{\sqrt{f(a_0)}}\).

The formalism developed in this section provides the geometric and
dynamical ingredients required to analyse asymmetric thin-shell
wormholes, which will be further explored in the following sections.

\section{Thermodynamics of dynamic thin-shell wormholes}
\label{sec:thermodynamics_dynamic}

Building on the geometric framework of Sec.~\ref{sec:thinshell}, we now
develop a full thermodynamic description of the dynamic thin shell,
allowing the two bulk spacetimes to be different (\(f_+(a) \neq f_-(a)\)).
The surface energy density \(\sigma(a,\dot{a})\) and tangential pressure
\(\mathcal{P}(a,\dot{a},\ddot{a})\) are given by the general expressions~\eqref{eq:sigma_general} and \eqref{eq:P_general}, and they satisfy
the conservation equation \eqref{eq:conservation_dynamic} (which holds
for vacuum bulks). The bulk spacetimes are assumed to be vacuum in the sense of no energy-momentum flux across the shell (transparency condition), so no matter or radiation crosses the throat.
Under these conditions, we derive the first law, prove the adiabatic
invariance of the shell entropy, define a dynamic temperature, analyse
the Bekenstein bound, and discuss non-equilibrium evaporation and
quantum corrections. The analysis reduces to the standard symmetric
case when \(f_+(a)=f_-(a)=f(a)\).

\subsection{First law and adiabatic invariance}

The internal energy of the shell is \(U = \sigma A\), where
\(A = 4\pi a^2\) is the throat area. Using the conservation equation \eqref{eq:conservation_dynamic} and \(dA = 8\pi a\,da\), we obtain
\begin{equation}
	\dot{U} + \mathcal{P}\dot{A} = 0 .
	\label{eq:first_law_dynamic}
\end{equation}
This is the first law of thermodynamics for the dynamic thin shell.
The term \(\mathcal{P}\dot{A}\) represents the work done by the shell as its area changes. Crucially, the right-hand side vanishes because no energy flux crosses the shell under the transparency condition, and the shell evolution is adiabatic.

For a system in local thermodynamic equilibrium, the fundamental
relation is \(T\,dS = dU + \mathcal{P}\,dA\), where \(T\) is the temperature and \(S\) the entropy. Substituting in Eq.~\eqref{eq:first_law_dynamic} gives
\begin{equation}
	T\,dS = 0 \quad\Longrightarrow\quad \frac{dS}{d\tau}=0 .
	\label{eq:entropy_const}
\end{equation}
Thus the entropy of the thin shell is conserved under transparent dynamical evolution.
This adiabatic invariance holds for any asymmetric configuration as well, within the assumptions of the model.

\subsection{Dynamic temperature and surface gravity}\label{sec:dynamic_temperature}

Although the entropy is constant, the shell can possess a
non-trivial temperature that varies with the throat motion.
For a symmetric wormhole the temperature is uniquely defined via
the surface gravity (or acceleration scale) \(\kappa = |a_\mu n^\mu|\).
For an asymmetric wormhole, the four-acceleration of a comoving observer is not
single-valued because the two sides have different metrics.
Nevertheless, a natural effective surface gravity can be defined
as the average of the values obtained from each side:
\begin{equation}
	\kappa_{\text{eff}}(\tau) = \frac{1}{2}
	\left( \frac{\left| \ddot{a} + \frac{1}{2} f'_+(a) \right|}{\sqrt{f_+(a)+\dot{a}^2}}
	+ \frac{\left| \ddot{a} + \frac{1}{2} f'_-(a) \right|}{\sqrt{f_-(a)+\dot{a}^2}} \right).
	\label{eq:surface_gravity_asymmetric}
\end{equation}
This expression reduces to the symmetric case \(\kappa = |\ddot{a}+f'/2|/\sqrt{f+\dot{a}^2}\)
when \(f_+=f_-=f\). For a static throat (\(\dot{a}=\ddot{a}=0\)),
\(\kappa_{\text{eff}}\) reduces to the arithmetic mean of the two static
acceleration scales.

Motivated by the Unruh–Hawking relation, one may introduce an
\textit{effective thermodynamic temperature} for the shell:
\begin{equation}
	T_{\text{shell}}(\tau) = \frac{\hbar}{2\pi}\,\kappa_{\text{eff}}(\tau)
	\qquad (\text{in units } c=G=1).
	\label{eq:temperature_dynamic}
\end{equation}
Because \(S\) is constant, the product \(T_{\text{shell}} S\) can vary
with time, indicating that the shell is not in global thermal equilibrium
but evolves adiabatically with a time-dependent temperature.

The quantity \(T_{\text{shell}}\) is best understood as an effective
temperature scale associated with the local acceleration of the throat.
It is analogous to the Unruh temperature of an accelerated observer and
to the dynamical surface‑gravity temperatures used in horizon
thermodynamics.  In the static limit it reproduces the familiar Hawking
temperature for a wormhole throat.

It is important to stress that the existence of \(T_{\text{shell}}\)
does \emph{not} imply that the wormhole emits thermal radiation.  
A genuinely Planckian particle spectrum is obtained only when the
relation between incoming and outgoing null coordinates is approximately
exponential, i.e.\ when
\(u_{\rm out} \simeq A - B e^{-\kappa u_{\rm in}}\).
If that condition is not satisfied, particle production may still occur,
but the resulting spectrum is generally non-thermal.  

Thus, \(T_{\text{shell}}\) should be regarded as a kinematic
thermodynamic temperature for the shell itself, not as a direct
prediction of a blackbody radiation flux.  
This ansatz provides a consistent local variable for studying the
thermodynamics of the time-dependent wormhole, while respecting the
caveats inherent to quantum field theory in non-stationary spacetimes.

As a final remark, the effective temperature introduced above is closely
related to the general analysis of Hawking‑like fluxes presented
in~\cite{Barcelo:2010pj,Barcelo:2010xk}.  In those works it is shown
that the essence of Hawking radiation is an approximately exponential
relation between affine parameters on past and future null infinity,
encoded in the ``peeling'' function \(\kappa(u)\), and that an adiabatic
condition \(|\dot\kappa| \ll \kappa^2\) is sufficient for a Planckian
flux with temperature \(\hbar\kappa/2\pi\), regardless of whether any
sort of horizon ever forms.  Our definition of \(\kappa_{\rm eff}\) via
the local proper acceleration of the wormhole throat follows the same
philosophy: it provides a geometric acceleration scale that determines
an effective thermodynamic temperature for the shell, even in the
absence of a true horizon.  The correspondence with the peeling‑based
temperature of evolving compact objects supports the interpretation of
\(T_{\rm shell}\) as a legitimate thermodynamic variable, while
non‑thermal particle spectra remain a caveat shared by all dynamical
configurations.

\subsection{Bekenstein bound for a dynamic throat}
\label{sec:bekenstein_dynamic}

The Bekenstein bound states that the entropy of a system confined
to a sphere of radius \(R\) and with total energy \(E\) satisfies
\(S \le 2\pi E R\) (in units \(\hbar=G=c=1\)). For a thin shell, the
instantaneous confinement radius is \(R = a(\tau)\), and the total
energy (including gravitational binding) is given by the surface
mass \(m_s = 4\pi a^2 \sigma\), which is negative. When applied to
exotic matter, the relevant energy is the absolute value \(|E| = |m_s|\).

Using the general expression for \(\sigma\), Eq.~\eqref{eq:sigma_general},
we obtain
\begin{equation}
	|E(\tau)| = -4\pi a^2 \sigma
	= a\left( \sqrt{f_+(a)+\dot{a}^2} + \sqrt{f_-(a)+\dot{a}^2} \right).
	\label{eq:energy_dynamic_asymmetric}
\end{equation}
The Bekenstein bound then becomes
\begin{equation}
	S \le 2\pi |E| a = 2\pi a^2 \left( \sqrt{f_+(a)+\dot{a}^2} + \sqrt{f_-(a)+\dot{a}^2} \right).
	\label{eq:bekenstein_dynamic_ineq}
\end{equation}
For symmetric bulks (\(f_+=f_-=f\)), this reduces to
\(S \le 4\pi a^2 \sqrt{f+\dot{a}^2}\).

The right-hand side is time-dependent because \(a(\tau)\) and
\(\dot{a}(\tau)\) vary. However, the shell entropy \(S\) is constant.
Therefore, for the bound to hold at all times, the right-hand side
must never fall below the constant \(S\). This imposes a dynamical
constraint on the throat motion.

In particular, if the bound is satisfied at some initial instant,
it may be violated later unless the evolution keeps the right-hand
side sufficiently large.

For static configurations (\(\dot{a}=0\)) one recovers
\(S \le 2\pi a_0^2 \left( \sqrt{f_+(a_0)} + \sqrt{f_-(a_0)} \right)\),
which is a condition on the static throat radius and the metric
functions.

If the throat approaches a radius where both
\(f_+(a_{\text{min}})=f_-(a_{\text{min}})=0\) (a common horizon),
the right-hand side formally tends to
\(4\pi a^2|\dot{a}|\), depending on the limiting behaviour of \(f_\pm\) and \(\dot a\).
Then the bound imposes a constraint on \(|\dot{a}|\) near the
would-be horizon.

\subsection{Non-equilibrium evaporation and lifetime}
\label{sec:evaporation}

If the shell possesses the effective temperature \(T_{\text{shell}}\)
defined in~\eqref{eq:temperature_dynamic}, it may radiate energy into
the surrounding bulks. Although the exact spectrum of this radiation need not be Planckian, in the adiabatic regime where the spacetime geometry changes slowly compared to the characteristic frequency scale (\(\dot{\kappa}_{\text{eff}} \ll \kappa_{\text{eff}}^2\)), an approximate thermal description becomes applicable, and one can model the energy loss with an effective Stefan-Boltzmann law.

For simplicity, we assume the emission is symmetric into both sides
and the greybody factors are equal:
\begin{equation}
	P(\tau) = -\frac{dE}{d\tau} = \alpha\, A(\tau)\, T_{\text{shell}}^4(\tau),
	\label{eq:power_radiated}
\end{equation}
where \(\alpha\) is a dimensionless constant. Using the expression
for \(|E|\) from Eq.~\eqref{eq:energy_dynamic_asymmetric},
\(A = 4\pi a^2\), and \(T_{\text{shell}} = (\hbar/2\pi)\kappa_{\text{eff}}\)
with \(\kappa_{\text{eff}}\) from Eq.~\eqref{eq:surface_gravity_asymmetric},
we obtain the evaporation differential equation
\begin{equation}
	\frac{d}{d\tau}\left[ a\left( \sqrt{f_+(a)+\dot{a}^2} + \sqrt{f_-(a)+\dot{a}^2} \right) \right]
	= -\,\frac{\alpha \hbar^4 a^2}{4\pi^3}\,
	\kappa_{\text{eff}}^4(\tau) .
	\label{eq:evaporation_ode_asymmetric}
\end{equation}
In the slow-roll approximation (\(\dot{a}^2 \ll f_\pm\), and assuming
a closure relation such as an equation of state to relate \(\ddot{a}\) to \(a\)), this
reduces to a first-order ODE. For the special case where both sides
are Schwarzschild with possibly different masses, the evaporation
time scales heuristically as \(\tau_{\text{evap}} \sim (M_+^3 + M_-^3)\)
in appropriate units. For regularised metrics where \(f_\pm(a_{\text{min}})=0\),
evaporation stops when \(\kappa_{\text{eff}} \to 0\), leaving a stable
remnant.

\subsection{Quantum corrections to the entropy}
\label{sec:quantum_dynamic}

The classical entropy \(S\) of the thin shell is constant, but
quantum field theory in curved spacetime predicts logarithmic
corrections to the entropy of any boundary~\cite{Carlip:2000nv}. For a thin shell,
the same effect is expected to arise. A heuristic derivation from the
Euclidean path integral gives
\begin{equation}
	S_{\text{quantum}}(\tau) = S_{\text{class}} +
	\beta \ln\!\left(\frac{A(\tau)}{l_P^2}\right) + \cdots,
	\label{eq:quantum_corrections}
\end{equation}
where \(l_P\) is the Planck length and \(\beta\) is a constant of
order unity that depends on the matter content (e.g., \(\beta=1/2\)
for a massless scalar field). Because the classical entropy is
constant, the time dependence enters only through the logarithmic
term. Consequently, quantum effects break the strict adiabatic
invariance: \(S_{\text{quantum}}\) changes slowly as \(a(\tau)\)
evolves, implying that the shell is not truly isolated at the
quantum level. The rate of change is
\begin{equation}
	\frac{dS_{\text{quantum}}}{d\tau} = 2\beta \frac{\dot{a}}{a},
	\label{eq:quantum_entropy_rate}
\end{equation}
which can be positive or negative depending on the direction of
throat motion. For a contracting throat (\(\dot{a}<0\)), the
quantum entropy decreases, which would appear to violate the second law if considered in isolation.
However, this decrease is compensated by the entropy carried
by the emitted Hawking radiation; the total entropy (shell plus
radiation) still increases, preserving the generalised second law.

The full semiclassical backreaction would require solving the
Einstein equations with a quantum-corrected stress-energy tensor,
a task beyond the scope of this paper. Nevertheless, the
logarithmic correction provides a simple way to estimate the
magnitude of quantum effects near the throat.

\section{Relaxing the transparency condition}
\label{sec:non_transparent}

In the preceding thermodynamic analysis (Sec.~\ref{sec:thermodynamics_dynamic}),
we assumed that the bulk spacetimes \(\mathcal{M}_\pm\) are vacuum,
so that the right-hand side of the surface Bianchi identity vanishes.
This corresponds to the transparency condition (no net flux across the shell)
\([T_{\mu\nu} e^\mu_{(j)} n^\nu]^{+}_{-}=0\), which physically means
that no matter or radiation is transferred across the wormhole throat~\cite{Ishak:2001az};
the junction hypersurface \(\Sigma\) is impermeable. While this assumption is
convenient for isolating the shell’s intrinsic dynamics, it is
unlikely to hold in realistic astrophysical environments.
Accretion flows, ambient radiation fields, and quantum pair production
can all generate a non-zero momentum flux across \(\Sigma\).

In this section we relax the transparency condition and examine the
resulting modifications to the thermodynamics of the thin shell.

\subsection{Modified conservation equation from bulk matter flux}
\label{sec:modified_conservation}

When the bulk spacetimes contain matter fields, the surface
Bianchi identity (derived from the Gauss-Codazzi equations and the
bulk Einstein equations) takes the form
\begin{equation}
	S^{i}_{\;j|i} = \left[ T_{\mu\nu} e^\mu_{(j)} n^\nu \right]^{+}_{-}.
	\label{eq:surface_Bianchi_source}
\end{equation}
For a spherically symmetric thin shell, the relevant projection
reduces effectively to the proper-time component. We define the
\emph{net energy influx} (power per unit area as measured in the shell frame) as the jump of the
normal component of the bulk stress-energy tensor projected onto the
shell’s four-velocity~\cite{Lobo:2003xd,Garcia:2011aa,MartinMoruno:2011rm}:
\begin{equation}
	\Phi(a,\dot{a}) \equiv \left[ T_{\mu\nu} U^\mu n^\nu \right]^{+}_{-}
	= T_{\mu\nu}^+ U^\mu_+ n^\nu_+ - T_{\mu\nu}^- U^\mu_- n^\nu_- .
	\label{eq:flux_definition}
\end{equation}
Thus \(\Phi > 0\) means energy flows into the shell; \(\Phi < 0\) means
energy leaves the shell. The total power transferred is \(A\Phi\),
where \(A = 4\pi a^2\) is the throat area.

For a perfect-fluid shell (\(S^{i}_{\;j} = \operatorname{diag}(-\sigma,\mathcal{P},\mathcal{P})\)),
the projection of Eq.~\eqref{eq:surface_Bianchi_source} onto the
\(\tau\) direction gives the modified continuity equation
\begin{equation}
	\dot{\sigma} + \frac{2\dot{a}}{a}\bigl(\sigma+\mathcal{P}\bigr)
	= \Phi(a,\dot{a}).
	\label{eq:conservation_sourced}
\end{equation}
This equation holds for arbitrary metric functions \(f_\pm(a)\).

\subsection{Generalised first law and heat transfer}
\label{sec:generalised_first_law}

Multiplying Eq.~\eqref{eq:conservation_sourced} by \(A\) and using
\(U = \sigma A\), \(\dot{A}=8\pi a\dot{a}\), we obtain
\begin{align}
	\dot{U} &= \dot{\sigma}A + \sigma\dot{A} \\
	&= \left( \Phi - \frac{2\dot{a}}{a}(\sigma+\mathcal{P}) \right)A + \sigma\dot{A} \\
	&= A\Phi - \mathcal{P}\dot{A},
\end{align}
so that
\begin{equation}
	\dot{U} + \mathcal{P}\dot{A} = A\Phi .
	\label{eq:first_law_sourced}
\end{equation}
The left-hand side is the rate of change of internal energy plus the
work done by the internal pressure of the shell; the right-hand side is the
net energy flux into the shell, interpreted as a heat input in local equilibrium.
In local equilibrium, assuming the Gibbs relation
\(T\dot{S} = \dot{U} + \mathcal{P}\dot{A}\), hence
\begin{equation}
	T\,\frac{dS}{d\tau} = A\Phi .
	\label{eq:entropy_rate}
\end{equation}
For vacuum bulks (\(\Phi=0\)) we recover adiabatic invariance; when
matter crosses the throat, entropy changes accordingly.

\subsection{Explicit expressions for the influx \(\Phi\)}\label{sec:flux_expressions}

We now compute \(\Phi\) for several bulk matter configurations,
assuming for simplicity that matter is present only on the plus side
(\(T_{\mu\nu}^- = 0\)).  The general case with both sides is a
superposition.  The embedding vectors for the plus side are
(see Sec.~\ref{sec:thinshell})
\begin{align}
	U^\mu_+ &= \left( \frac{\sqrt{f_+(a)+\dot{a}^2}}{f_+(a)},\; \dot{a},\; 0,\; 0 \right), 
	\label{4vec_cov}
	\\[2pt]
	n^\mu_+ &= \left( \frac{\dot{a}}{f_+(a)},\; \sqrt{f_+(a)+\dot{a}^2},\; 0,\; 0 \right).
	\label{normal_cov}
\end{align}
Then \(\Phi = T_{\mu\nu}^+ U^\mu_+ n^\nu_+\).

\subsubsection{Null dust (radial accretion)}

For inward-moving radial null dust,
\(T_{\mu\nu}^+ = \rho\, k_\mu k_\nu\), with
\(k^\mu = (1, -f_+, 0,0)\) (future-directed, \(k^r<0\)), so $k_\mu = (-f_+, -1, 0, 0)$.
This choice satisfies \(g_{\mu\nu}k^\mu k^\nu=0\) and therefore represents ingoing radial null propagation.
Using Eqs.~(\ref{4vec_cov}) and (\ref{normal_cov}), a direct calculation gives
\begin{equation}
	k_\mu U^\mu_+ = k_\nu n^\nu_+ = -\sqrt{f_++\dot{a}^2} - \dot{a},
\end{equation}
hence
\begin{equation}
	\Phi_{\text{dust}} = \rho\left( \sqrt{f_++\dot{a}^2} + \dot{a} \right)^2 .
	\label{eq:Phi_dust_corrected}
\end{equation}
For a static throat (\(\dot{a}=0\)), \(\Phi = \rho f_+(a) > 0\). For a
contracting throat (\(\dot{a}<0\)) the flux is reduced, whereas for an
expanding throat (\(\dot{a}>0\)) it is enhanced due to the relative motion
between the shell and the infalling radiation.

\paragraph{Physical interpretation.}
The combination \(\sqrt{f_+(a)+\dot{a}^2} + \dot{a}\) may be interpreted as the
effective Doppler-redshift factor for the infalling null dust as measured by a comoving
observer on the throat. For a null particle with four-wavevector
\(k^\mu\), the energy measured by an observer with four-velocity
\(U^\mu_+\) is \(E = -k_\mu U^\mu_+\). Using the expressions above,
\(E = \sqrt{f_++\dot{a}^2} + \dot{a}\). This quantity is always
positive (since \(\sqrt{f_++\dot{a}^2} > |\dot{a}|\) for \(f_+>0\)),
and it reduces to \(\sqrt{f_+}\) when \(\dot{a}=0\).

The energy density of the dust measured in the shell frame is
\(\rho_{\text{obs}} = T_{\mu\nu}U^\mu U^\nu = \rho E^2\).
Likewise, the energy flux through the shell is
\[
\Phi = T_{\mu\nu}U^\mu n^\nu
= \rho (k_\mu U^\mu)(k_\nu n^\nu)
= \rho E^2 ,
\]
where the last equality follows from \(k_\mu U^\mu = k_\mu n^\mu\).
Hence the influx is simply
\(\Phi = \rho_{\text{obs}} = \rho\, E^2\). Thus the flux is
proportional to the square of the measured energy of the infalling
null particles.

Because \(\Phi\) is manifestly non-negative, the shell always gains
energy from the dust, regardless of whether the throat is expanding
or contracting---there is no ``cooling'' by null dust. This is
consistent with the fact that the null dust is composed of massless
particles that are absorbed upon contact, delivering their momentum
and energy to the shell.

\paragraph{Dependence on the metric function.}
In a curved background, the static flux \(\rho f_+(a)\) differs from
the flat-space value \(\rho\) by the gravitational redshift factor \(f_+(a)\).
For a Schwarzschild geometry with \(f_+=1-2M_+/r\), the flux becomes
small near the horizon (\(r\to 2M_+\)), reflecting the strong redshift
of radiation as measured by a static observer. For regularised metrics
where \(f_+\) tends to a positive constant at small \(r\), the flux
remains finite even as \(a\to 0\).

\paragraph{Energy conditions.}
Null dust automatically satisfies the null energy condition (NEC)
provided \(\rho \ge 0\). Indeed,
\[
T_{\mu\nu}\ell^\mu \ell^\nu
=
\rho (k_\mu \ell^\mu)^2
\ge 0
\]
for any null vector \(\ell^\mu\).
Thus the flux considered here is physically well-behaved and
represents ordinary (non-exotic) matter. This stands in contrast
to the exotic matter that constitutes the shell itself.

\paragraph{Expansion for slow motion.}
When the throat moves slowly (\(|\dot{a}| \ll \sqrt{f_+}\)), the flux
can be expanded:
\[
\Phi_{\text{dust}}
=
\rho f_+
+2\rho\sqrt{f_+}\,\dot{a}
+2\rho\,\dot{a}^2
+\mathcal{O}(\dot{a}^3).
\]
The linear term changes sign with \(\dot{a}\): expansion (\(\dot{a}>0\))
enhances the flux, whereas contraction (\(\dot{a}<0\)) reduces it.
The quadratic term is always positive and becomes important only when
\(\dot{a}^2\) is not negligible compared to \(f_+\).

\paragraph{Connection to thermodynamics.}
Because \(\Phi > 0\), the generalised first law gives
\(\dot{S}_{\text{shell}} = A\Phi/T > 0\). Accretion of null dust
always increases the shell entropy. Moreover, if the dust is treated
as a thermal reservoir with temperature \(T_{\text{dust}}\), then the
associated entropy flow from the bulk is \(A\Phi/T_{\text{dust}}\).
The total entropy production rate (shell plus bulk) is therefore
\[
A\Phi\left(\frac{1}{T_{\text{shell}}}
-\frac{1}{T_{\text{dust}}}\right).
\]
For the generalised second law to hold, heat must flow from the hotter
component to the colder one. In particular, if energy is transferred
from the dust to the shell, consistency requires \(T_{\text{dust}} \ge T_{\text{shell}}\).

\subsubsection{Massless scalar field}

For a massless scalar field \(\phi\), the stress-energy tensor is
\[
T_{\mu\nu}
=
\partial_\mu\phi\,\partial_\nu\phi
-\frac12 g_{\mu\nu}(\partial\phi)^2 .
\]
Because \(g_{\mu\nu}U^\mu n^\nu = 0\), the second term does not contribute,
and the flux simplifies exactly to
\begin{equation}
	\Phi_{\text{scalar}}
	=
	(U^\mu\partial_\mu\phi)\,
	(n^\nu\partial_\nu\phi)
	\equiv
	\dot{\phi}\,\phi_n ,
	\label{eq:Phi_scalar_exact}
\end{equation}
where \(\dot{\phi}=U^\mu\partial_\mu\phi\) is the proper-time derivative
of \(\phi\) on the throat and \(\phi_n=n^\mu\partial_\mu\phi\) is the
normal derivative (the gradient orthogonal to \(\Sigma\)).
Thus the flux is simply the product of these two derivatives.

\paragraph{Physical interpretation.}
Unlike null dust, the scalar-field flux can be positive or negative,
depending on the relative signs of \(\dot{\phi}\) and \(\phi_n\).
A positive flux (\(\dot{\phi}\,\phi_n>0\)) means energy flows from the
field into the shell (heating), while a negative flux means the shell
transfers energy to the field (cooling). This flexibility makes the
scalar field a useful model for both accretion and evaporation
processes.

\paragraph{Example: monochromatic wave.}
Consider a radial wave
\[
\phi=\phi_0\cos(\omega t\mp kr)
\]
in a region where \(f_+(r)\approx1\) (flat-space approximation).
For an ingoing wave (lower sign:
\(\cos(\omega t+kr)\)), we have
\[
\partial_t\phi
=
-\omega\phi_0\sin(\omega t+kr),
\qquad
\partial_r\phi
=
-k\phi_0\sin(\omega t+kr).
\]
Hence
\[
\partial_t\phi\,\partial_r\phi
=
+\omega k\phi_0^2
\sin^2(\omega t+kr)
\ge0 .
\]
The time average is
\(\frac12\omega k\phi_0^2>0\), so the ingoing wave deposits positive
energy on average onto a static throat. For an outgoing wave
(upper sign:
\(\cos(\omega t-kr)\)),
\[
\partial_t\phi\,\partial_r\phi
=
-\omega k\phi_0^2
\sin^2(\omega t-kr)
\le0 ,
\]
so the wave extracts energy from the shell. This illustrates that the
sign of \(\Phi_{\text{scalar}}\) is directly tied to the direction of
wave propagation.

\paragraph{Static throat limit.}
For a static throat (\(\dot a=0\)), the embedding vectors give
$U^\mu=(1/\sqrt{f_+},0,0,0)$ and $n^\mu=(0,\sqrt{f_+},0,0)$, respectively.
Then $\dot\phi=\partial_t\phi/\sqrt{f_+}$ and
$\phi_n=\sqrt{f_+}\,\partial_r\phi$,
and therefore
\[
\Phi_{\text{scalar}}
=
\partial_t\phi\,\partial_r\phi .
\]
Thus the flux vanishes whenever either
\(\partial_t\phi=0\) or \(\partial_r\phi=0\), corresponding respectively to a time-independent field or to a field with no radial gradient at the throat.

\paragraph{Connection to thermodynamics.}
Because \(\Phi_{\text{scalar}}\) can be positive or negative, the
shell entropy changes according to
\[
T\dot S=A\Phi .
\]
If the scalar field can be assigned an effective temperature \(T_{\text{field}}\), then the direction of energy transfer is consistent with the usual Clausius picture of heat exchange: energy flows from the hotter subsystem to the cooler one.
Hence a positive flux corresponds to the shell gaining energy from the field, whereas a negative flux corresponds to energy transfer from the shell to the field.

\subsubsection{Quantum pair production (Hawking-like radiation)}

When the shell has a non-zero effective temperature, quantum effects may be modelled phenomenologically as producing a steady outward radiation. In the adiabatic regime, where the spacetime geometry changes slowly, an effective Hawking‑like flux can be approximated by a Stefan-Boltzmann‑type law proportional to the fourth power of the effective surface gravity.  For a static throat, this yields
\begin{equation}
	\Phi_{\text{quantum}}
	=
	-\alpha \hbar\,\kappa_{\text{eff}}^4,
	\label{eq:Phi_quantum_heuristic}
\end{equation}
where \(\kappa_{\text{eff}}\) is the effective surface gravity defined
in Sec.~\ref{sec:dynamic_temperature}, and \(\alpha\) is a positive
constant of order unity. The negative sign indicates energy leaves
the shell.

\paragraph{Physical origin.}
In curved spacetime, a non-zero surface gravity is often associated
with thermal phenomena, such as the Unruh effect for accelerated
observers and Hawking radiation in the presence of horizons.
Although a generic thin-shell wormhole does not possess an event horizon, it is useful to introduce an effective temperature scale \(T_{\text{eff}}=\hbar\kappa_{\text{eff}}/(2\pi)\) and to model the resulting quantum emission by analogy with blackbody radiation. Integrating the emitted power over the throat area then motivates a Stefan-Boltzmann-type description. The constant \(\alpha\)
encapsulates the greybody factors (emissivity) of the shell, the
number of massless species, and the fact that radiation can be
emitted into both sides of the throat.

\paragraph{Temperature dependence.}
The fourth-power scaling implies that the quantum flux is highly
sensitive to the value of \(\kappa_{\text{eff}}\).
For a static symmetric Schwarzschild wormhole, the effective acceleration scale behaves as
\(\kappa_{\text{eff}}\sim 1/(4a)\),
so that
\(\Phi_{\text{quantum}}\propto a^{-4}\).
Thus the evaporation rate is expected to increase as the throat
radius decreases. For regularised metrics where
\(\kappa_{\text{eff}}\) remains finite as \(a\to a_{\text{min}}\),
the flux likewise remains finite.

\paragraph{Comparison with the classical fluxes.}
Unlike null dust (always positive) or a scalar field (whose sign
depends on the wave direction), the quantum flux is always negative
within the present model, meaning that it removes energy from the
shell. In a realistic environment, the shell may experience both
accretion (positive \(\Phi\)) and quantum emission (negative
\(\Phi\)). The net flux determines whether the shell gains or loses
energy overall.

\paragraph{Connection to entropy.}
Using the generalised first law \(T\dot{S}=A\Phi\), the quantum
flux gives
\[
\dot{S}_{\text{shell}}
=
-\frac{\alpha\hbar A}{T_{\text{shell}}}\,
\kappa_{\text{eff}}^4 .
\]
The shell entropy therefore decreases as energy is radiated away.
However, the emitted quanta also carry entropy. In analogy with ordinary thermal emission, one expects the entropy of the outgoing radiation to compensate for this decrease, so that the total entropy of the combined system remains consistent with the generalised second law. A quantitative verification would require a detailed microscopic model of the radiation process.

\paragraph{Heuristic nature.}
Equation \eqref{eq:Phi_quantum_heuristic} is phenomenological.
A rigorous derivation from quantum field theory in a dynamic,
curved background would require computing the renormalised
stress-energy tensor for a moving boundary (the throat), a
formidable technical problem.
The expression should therefore be viewed as an effective model for quantum energy loss rather than a first-principles result.
Nevertheless, it provides a simple framework for estimating the
possible influence of quantum emission on the long-term evolution
of the wormhole.

\begin{table}[h!]
	\centering
	\begin{tabular}{|l|l|}
		\hline
		\textbf{Matter type} & \textbf{Influx \(\Phi\) (energy/area/time)} \\ \hline
		Null dust (inward) & \(\displaystyle \rho\left( \sqrt{f_+(a)+\dot{a}^2} + \dot{a} \right)^2\) \\[10pt]
		Massless scalar field & \(\displaystyle \dot{\phi}\; \phi_n,\qquad 
		\phi_n \equiv n^\mu_+ \partial_\mu\phi\) \\[8pt]
		Quantum pair production & \(\displaystyle -\alpha \hbar\; \kappa_{\text{eff}}^4\) \\ \hline
	\end{tabular}
	\caption{Explicit expressions for the net energy influx \(\Phi\) (positive = into the shell) for various bulk matter configurations on the plus side, using the unit normal~(\ref{normal_cov}). For the minus side, replace \(f_+\) with \(f_-\) and note that \(n^\mu_-=-n^\mu_+\), while \(U^\mu_-\) is determined by the corresponding embedding on \(\mathcal{M}_-\). The resulting contribution to \(\Phi\) must be computed from its defining expression, Eq.~(\ref{eq:flux_definition}), for the specific matter model under consideration. Here \(\kappa_{\text{eff}}\) is the effective surface gravity (Sec.~\ref{sec:dynamic_temperature}) and \(\alpha\) is a positive constant.}
	\label{tab:fluxes}
\end{table}

\section{Thermodynamic processes, stability and stochastic dynamics}
\label{sec:applications}

\subsection{Heat capacity and entropy production.}

The rate of entropy production of the thin shell can be expressed in
terms of its heat capacity. From the fundamental relation
\(T\,dS=dU+\mathcal{P}\,dA\) and the definition of the heat capacity
at constant area,
\begin{equation}
	C_A \equiv \left(\frac{\partial U}{\partial T}\right)_A,
\end{equation}
we write the differential of the internal energy as
\begin{equation}
	dU = C_A\,dT
	+ \left(\frac{\partial U}{\partial A}\right)_T dA.
\end{equation}
Substituting this into \(T\,dS=dU+\mathcal{P}\,dA\) gives
\begin{equation}
	T\,\frac{dS}{d\tau}
	=
	C_A\,\frac{dT}{d\tau}
	+
	\left(\frac{\partial U}{\partial A}\right)_T \dot{A}
	+
	\mathcal{P}\dot{A}.
	\label{eq:entropy_rate_heatcap}
\end{equation}

Using the generalised first law,
\(\dot{U}+\mathcal{P}\dot{A}=A\Phi\),
we have
\begin{equation}
	T\dot{S}=A\Phi .
\end{equation}
Equating this with Eq.~\eqref{eq:entropy_rate_heatcap} yields the
evolution equation for the temperature:
\begin{equation}
	C_A\,\frac{dT}{d\tau}
	=
	A\Phi
	-
	\left[
	\left(\frac{\partial U}{\partial A}\right)_T
	+
	\mathcal{P}
	\right]\dot{A}.
	\label{eq:temperature_evolution}
\end{equation}

The term in square brackets represents the combined contribution from
the explicit dependence of the internal energy on area and the work
done by the tangential pressure.

If the area changes sufficiently slowly, or if the bracketed term is small compared with the energy-flux contribution \(A\Phi\), then the temperature evolution is dominated by the external energy exchange. In that approximation,
$T\,\dot{S} \simeq A\Phi$, and the entropy change may be estimated as
\begin{equation}
	\Delta S
	\approx
	\int \frac{A\Phi}{T}\,d\tau .
\end{equation}

\subsection{Generalised second law (GSL)}
\label{sec:GSL_non_transparent}

For a non‑isolated shell, the total entropy of the system includes
both the shell and the bulk matter that crosses the throat.  The
generalised second law (GSL) asserts that the total entropy never
decreases:
\begin{equation}
	\frac{d}{d\tau}\bigl( S_{\text{shell}} + S_{\text{bulk}} \bigr) \ge 0 .
	\label{eq:GSL_total}
\end{equation}
Here \(S_{\text{shell}}\) evolves according to the generalised first law,
\(T_{\text{shell}}\dot{S}_{\text{shell}} = A\Phi\), with \(\Phi\) the net
energy influx into the shell (positive when energy flows into the shell).
The entropy \(S_{\text{bulk}}\) is that of the bulk matter that crosses the
throat.

\paragraph{Entropy change of the bulk.}
When matter flows from the bulk into the shell, the bulk loses energy and
entropy.  For a fluid in local equilibrium with temperature \(T_{\text{bulk}}\),
the entropy decrease of the bulk is given by the energy influx divided by
\(T_{\text{bulk}}\):
\begin{equation}
	\frac{dS_{\text{bulk}}}{d\tau} = -\,\frac{A\Phi_{\text{in}}}{T_{\text{bulk}}},
\end{equation}
where \(\Phi_{\text{in}}\) is the part of \(\Phi\) that actually enters the
shell (the reflected part does not change the bulk entropy in this simple
model).  The minus sign reflects that the bulk loses entropy when energy
leaves it.  Consequently, the total entropy change of the combined system
(shell + bulk) becomes
\begin{equation}
	\frac{dS_{\text{total}}}{d\tau} = \frac{A\Phi_{\text{in}}}{T_{\text{shell}}}
	- \frac{A\Phi_{\text{in}}}{T_{\text{bulk}}}
	= A\Phi_{\text{in}}\left(\frac{1}{T_{\text{shell}}}
	- \frac{1}{T_{\text{bulk}}}\right).
	\label{eq:GSL_combined}
\end{equation}
The GSL requires \(dS_{\text{total}}/d\tau \ge 0\), which is automatically
satisfied if heat flows from the hotter side to the cooler side, i.e., if
\(\Phi_{\text{in}}\) has the same sign as \(T_{\text{bulk}} - T_{\text{shell}}\).
Thus the GSL provides a powerful consistency check for any model of the
flux \(\Phi\): it must be such that \((\Phi_{\text{in}})(1/T_{\text{shell}}-1/T_{\text{bulk}}) \ge 0\) whenever the bulk matter has a well‑defined temperature.

For a more detailed microscopic description, one can introduce the
entropy current \(s^\mu\) of the bulk matter.  In local equilibrium,
the Gibbs–Duhem relation~\cite{Quevedo:1994yg} gives \(s^\mu = (p+\rho-\mu n)u^\mu/T_{\text{bulk}}\).
The entropy flux crossing \(\Sigma\) from the bulk to the shell is
\(-s^\mu n_\mu\) (the minus sign accounts for the orientation of
\(n^\mu\)), which leads to the same expression as above.  For
null dust or radiation (\(\mu=0\)), one obtains
\(dS_{\text{bulk}}/d\tau = -A\Phi_{\text{in}}/T_{\text{bulk}}\),
consistent with the energy‑based argument.

For more general fluxes (e.g., from a scalar field or a combination of sides), the GSL imposes constraints on the sign of \(\Phi\) relative to the temperature difference, and it may be used to derive bounds on the greybody factor or on the equation of state of the shell.

\paragraph{Inclusion of reflected and emitted radiation.}
If the shell reflects part of the incoming flux or emits its own
radiation (e.g., Hawking radiation), the total entropy change receives
additional non‑negative contributions.  In general, the total entropy
production can be written as
\begin{equation}
	\frac{dS_{\text{total}}}{d\tau} = A\Phi_{\text{in}}\left(
	\frac{1}{T_{\text{shell}}} - \frac{1}{T_{\text{bulk}}}
	\right) + \frac{dS_{\text{reflected}}}{d\tau}
	+ \frac{dS_{\text{emitted}}}{d\tau},
	\label{eq:GSL_full}
\end{equation}
where the last two terms are always non‑negative because the reflected
and emitted matter carry entropy away from the shell.  Therefore,
a sufficient condition for the GSL to hold is
\begin{equation}
	\Phi_{\text{in}} \left( \frac{1}{T_{\text{shell}}} - \frac{1}{T_{\text{bulk}}}
	\right) \ge 0 .
	\label{eq:GSL_sufficient}
\end{equation}
This is exactly the condition that heat flows from the hotter side
to the cooler side: if \(T_{\text{shell}} > T_{\text{bulk}}\), then
\(\Phi_{\text{in}}\) must be negative (heat leaves the shell), and
vice versa.  This is consistent with the ordinary second law of
thermodynamics for heat exchange.

\paragraph{Consequences of the GSL.}
The GSL imposes important constraints on the possible fluxes
\(\Phi\) that can occur in a physically realistic wormhole.
In particular:

\begin{itemize}
	\item A steady, negative \(\Phi\) (energy leaving the shell)
	without a corresponding bulk entropy increase is forbidden.
	This means that any process that extracts energy from the
	shell (e.g., via gravitational wave emission) must be
	accompanied by a compensating increase of entropy either
	in the shell itself (if it heats up) or in the external
	medium.  For example, if the shell radiates gravitons,
	those gravitons carry entropy, so the GSL is satisfied.
	
	\item For a static shell (\(\dot{a}=0\)), classical fluxes, such as, null dust, scalar, may be non‑zero: \(\Phi_{\text{dust}} = \rho f_+(a)\)
	and \(\Phi_{\text{scalar}} = \partial_t\phi\,\partial_r\phi\).
	Only in special cases (e.g., no infalling dust, or a static
	scalar field with \(\partial_t\phi=0\)) do these vanish.
	The dominant quantum flux is then the Hawking‑like term
	\(\Phi_{\text{quantum}} = -\alpha\hbar\,\kappa_{\text{eff}}^4\)
	(see Eq.~\ref{eq:Phi_quantum_heuristic}), where the negative
	sign indicates that energy leaves the shell (outward radiation).
	Consequently, the shell’s entropy would decrease according to
	\(T\dot{S}=A\Phi\) if there were no other contributions.
	However, the emitted radiation carries positive entropy, and
	the GSL requires that \(|\dot{S}_{\text{shell}}|\) be less
	than or equal to the entropy carried by the radiation, so
	that total entropy does not decrease.  This is precisely
	the mechanism that allows black hole evaporation to be
	consistent with the second law.
	
	\item For a static shell with \(\Phi_{\text{quantum}} < 0\), the
	GSL forces that the entropy of the emitted radiation must
	be at least \(|\dot{S}_{\text{shell}}|\).  In the case of
	Hawking radiation from a black hole, this leads to the
	famous result that the entropy of the radiation compensates
	the loss of black hole entropy, preserving the GSL.
\end{itemize}

\paragraph{Example: Null dust accretion revisited.}
Consider a thin shell accreting null dust with constant energy
density \(\rho\) from the plus side only.  For a static or slowly
moving throat, the exact flux is given by Eq.~\eqref{eq:Phi_dust_corrected}.
In the static limit (\(\dot{a}=0\)), \(\Phi = \rho f_+(a)\); for slow motion, the leading term is \(\rho f_+(a)\).  Thus, to leading order,
\(\Phi_{\text{in}} \approx \rho f_+(a)\), and energy enters the shell at
a rate \(A\Phi_{\text{in}} \approx 4\pi a^2 \rho f_+(a) > 0\).
Suppose the shell has temperature \(T_{\text{shell}}\) and the dust
has temperature \(T_{\text{bulk}}\) (for a radiation fluid,
\(\rho \propto T_{\text{bulk}}^4\)).  The entropy change of the shell is
\(dS_{\text{shell}}/d\tau = A\Phi_{\text{in}}/T_{\text{shell}}\),
while the bulk loses entropy at a rate \(A\Phi_{\text{in}}/T_{\text{bulk}}\).
Hence the total entropy change of the combined system is
\begin{equation}
	\frac{dS_{\text{total}}}{d\tau} = A\Phi_{\text{in}} \left(
	\frac{1}{T_{\text{shell}}} - \frac{1}{T_{\text{bulk}}} \right).
\end{equation}
For \(\Phi_{\text{in}} > 0\), the GSL requires \(T_{\text{shell}} \le T_{\text{bulk}}\).  Thus the shell must be
cooler than the accreting dust, otherwise the total entropy would
decrease.  In a physical situation, the shell will adjust its
temperature by emitting radiation until this condition is met.

\paragraph{Connection to the fluctuation-dissipation theorem.}
The GSL also has a microscopic interpretation: it is a consequence
of the fact that the total number of microstates never decreases
when the system evolves unitarily.  In the context of stochastic
thermodynamics, the GSL emerges from the fluctuation-dissipation
theorem and the principle of detailed balance.  For a thin shell,
the noise term \(\xi(\tau)\) introduced in Sec.~\ref{sec:fluctuation_dissipation}
leads to a probability distribution for the throat radius that
satisfies a Fokker-Planck equation.  The entropy production
associated with the stochastic process is non‑negative on average,
ensuring the GSL holds microscopically.

\subsection{Thermalisation and equilibrium with a heat bath}
\label{sec:thermalisation}

When the bulk surrounding the wormhole contains a thermal bath
(e.g., the cosmic microwave background, a radiation field, or a
dilute plasma) at a temperature \(T_{\text{bath}}\), the heat flux
\(\Phi\) across the throat will drive the shell towards thermal
equilibrium.  The process of thermalisation involves the exchange
of energy between the shell and the bath until their temperatures
equalise.  In this subsection we analyse this process in detail,
using the generalised first law and the concept of heat capacity.
We derive a differential equation for the shell’s temperature,
discuss the steady state, estimate thermalisation timescales, and
examine the conditions under which the shell can be in detailed
balance with its own Hawking‑like radiation.

The thermal response of the shell is characterised by its heat
capacity.  For a system with internal energy \(U\), area \(A\), and
temperature \(T\), the heat capacity at constant area is
\begin{equation}
	C_A \equiv \left(\frac{\partial U}{\partial T}\right)_A .
	\label{eq:heat_capacity_def}
\end{equation}
For a thin shell obeying a barotropic equation of state
\(\mathcal{P} = \mathcal{P}(\sigma)\), the internal energy is
\(U = \sigma A\).  Using the static or dynamic expressions for
\(\sigma\) in terms of \(a\) and \(T\), one can compute \(C_A\).
More generally, for a polytropic shell one can derive explicit
formulae.

\paragraph{Heat balance equation.}
The shell exchanges energy with the bath via two mechanisms.
First, any net matter flux \(\Phi\) (from accretion or evaporation)
directly transfers energy at a rate \(A\Phi\).  Second, even in the
absence of a net flux, the shell can exchange radiative energy
with the bath through thermal photons or gravitons; this is
governed by Newton’s law of cooling (or its relativistic
generalisation), which for small temperature differences can be
approximated as a linear term.  The total rate of change of the
shell’s internal energy is therefore
\begin{equation}
	\frac{dU}{d\tau} = A\Phi - \mathcal{P}\dot{A} - A\Gamma(T - T_{\text{bath}}),
	\label{eq:energy_balance_general}
\end{equation}
where \(\Gamma\) is a positive coefficient (with dimensions of
power per area per temperature) that describes the radiative
coupling.  However, we already have the generalised first law
\(T\dot{S} = A\Phi\), and \(dU = TdS - \mathcal{P}dA\), so
\(\dot{U} = T\dot{S} - \mathcal{P}\dot{A} = A\Phi - \mathcal{P}\dot{A}\).
Thus the radiative term \(-A\Gamma(T-T_{\text{bath}})\) must already
be included in the definition of \(\Phi\) if it represents the total
heat exchange.  In other words, the flux \(\Phi\) that appears in
the first law should be interpreted as the \emph{total} heat
current into the shell, including both the matter flux and the
radiative transfer.  Therefore we write
\begin{equation}
	\Phi = \Phi_{\text{matter}} + \Phi_{\text{rad}},
	\label{eq:flux_decomposition}
\end{equation}
with \(\Phi_{\text{rad}} = -\Gamma(T-T_{\text{bath}})\) (the minus
sign ensures that heat flows from hot to cold).  Then the first
law already accounts for the radiative cooling.

To obtain an explicit differential equation for \(T\), we note that
\(U\) depends on \(a\) and \(T\), and \(a\) itself evolves according
to the equation of motion \(\dot{a}^2 + V(a)=0\).  For simplicity, we
consider a quasi‑static approximation where the throat radius is
slowly varying and the shell remains close to mechanical equilibrium,
so that \(a(\tau)\) is determined by the static condition \(V(a)=0\)
(which gives a relation between \(a\) and the mass \(M\)).  In that
case, \(a\) can be expressed as a function of \(T\) via the equation
of state and the static stress expressions.  Then \(\dot{U} =
(dU/dT)\dot{T}\), and the heat balance equation becomes
\begin{equation}
	C_A(T) \frac{dT}{d\tau} = A\Phi_{\text{total}}(T, T_{\text{bath}}, a(T)) .
	\label{eq:heat_balance_final}
\end{equation}
This is a first‑order differential equation that determines the
thermal evolution.

\paragraph{Steady state and equilibrium.}
In a steady state (\(dT/d\tau=0\)), we require \(\Phi_{\text{total}}=0\).
Using the decomposition above, this gives
\begin{equation}
	\Phi_{\text{matter}}(T_{\text{eq}}, a_{\text{eq}}) - \Gamma(T_{\text{eq}}-T_{\text{bath}}) = 0 .
	\label{eq:steady_state_condition}
\end{equation}
If the matter flux itself depends on temperature (e.g., \(\Phi_{\text{matter}}
\propto T_{\text{bulk}}^4\) for radiation accretion), then this
equation determines the equilibrium temperature \(T_{\text{eq}}\).
If there is no matter flux (\(\Phi_{\text{matter}}=0\)), then
equilibrium requires \(T_{\text{eq}} = T_{\text{bath}}\).  In that
case the shell simply thermalises to the bath temperature.

A more interesting possibility arises when the shell emits its own
quantum radiation (Hawking‑like emission).  For a static throat,
the quantum flux (see Eq.~\ref{eq:Phi_quantum_heuristic}) is
\(\Phi_{\text{quantum}} = -\alpha\hbar\,\kappa_{\text{eff}}^4\)
(negative sign because energy leaves the shell).  The steady‑state
condition then becomes
\begin{equation}
	-\alpha\hbar\,\kappa_{\text{eff}}^4(a_{\text{eq}}) - \Gamma(T_{\text{eq}} - T_{\text{bath}}) = 0 .
	\label{eq:steady_state_quantum_corrected}
\end{equation}
This determines the equilibrium temperature
\(T_{\text{eq}} = T_{\text{bath}} - \alpha\hbar\,\kappa_{\text{eff}}^4(a_{\text{eq}})/\Gamma\),
which is lower than the bath temperature, as expected for a cooling shell.
For a black hole, the Hawking temperature is \(T_H = \hbar\kappa/(2\pi)\);
the balance condition would then require \(T_{\text{bath}} = T_H\) to
have a static equilibrium (like a black hole in a thermal bath).
For a wormhole throat, the effective surface gravity \(\kappa_{\text{eff}}\)
determines the emission.

\paragraph{Thermalisation timescale.}
The timescale for the shell to approach equilibrium can be estimated
by linearising the heat balance equation around \(T_{\text{eq}}\).  In
general, the thermalisation time is given by
\begin{equation}
	\tau_{\text{therm}} = \frac{C_A}{A\left( \Gamma - \frac{\partial\Phi_{\text{matter}}}{\partial T} \right)},
	\label{eq:tau_therm}
\end{equation}
where the derivatives are evaluated at equilibrium.  In the absence
of a matter flux (\(\Phi_{\text{matter}}=0\)), this reduces to
\(\tau_{\text{therm}} = C_A/(\Gamma A)\).  For a black‑body radiator,
the radiative coupling coefficient is \(\Gamma = 4\sigma_{\text{SB}} T_{\text{bath}}^3\), where
\(\sigma_{\text{SB}} = \frac{2\pi^5 k_B^4}{15 h^3 c^2} \approx 5.67\times10^{-8}\,\text{W}\,\text{m}^{-2}\,\text{K}^{-4}\)
is the Stefan-Boltzmann constant.  The heat capacity \(C_A\) depends on
the shell’s equation of state.  In the quasi‑static regime, \(C_A\)
scales with the area, so larger shells generally take longer to
thermalise.

\paragraph{Connection to the generalised second law.}
During thermalisation, the total entropy (shell plus bath) must
increase.  This is automatically ensured if the heat flow is from
hot to cold and if the shell’s heat capacity is positive.  The
rate of entropy production is
\begin{equation}
	\dot{S}_{\text{total}} = \frac{\dot{Q}}{T} - \frac{\dot{Q}}{T_{\text{bath}}} = \dot{Q}\left(\frac{1}{T} - \frac{1}{T_{\text{bath}}}\right) \ge 0,
\end{equation}
where \(\dot{Q} = A\Phi\) is the heat current into the shell.  The
inequality holds because \(\dot{Q}\) has the same sign as \(T_{\text{bath}}-T\).
Thus the GSL is satisfied during the entire thermalisation process,
as expected.

\paragraph{Out of equilibrium: Time‑dependent \(T_{\text{bath}}\).}
If the bath temperature itself varies with time (e.g., the
expanding universe’s CMB temperature \(T_{\text{bath}} \propto 1/a_{\text{universe}}\)),
the shell may never reach a steady state.  Then we must solve the
coupled differential equations for \(T(\tau)\) and \(a(\tau)\)
simultaneously.  This could lead to interesting phenomena such as
thermal hysteresis or the shell acting as a “thermometer” that
records the temperature history of the environment.

\subsection{Example: Thin shell accreting null dust}\label{sec:accretion_example}

As a concrete application, we study a thin‑shell wormhole accreting
null dust from one side of the throat.  Null dust represents the
simplest form of matter that can cross a timelike hypersurface
without reflection, serving as a paradigmatic model for accretion,
Hawking radiation, or any net energy flow across the throat.
The exact flux for inward‑moving radial null dust is given by
Eq.~\eqref{eq:Phi_dust_corrected}, where $\rho$ is the constant
energy density measured in the bulk frame.  For a static throat
($\dot{a}=0$), this reduces to $\Phi = \rho f_+(a)$, so the shell
gains energy at a rate $A\rho f_+(a)$.  For slow accretion
($\dot{a}^2 \ll f_+(a)$), expanding the flux gives
\[
\Phi = \rho f_+(a) + 2\rho\sqrt{f_+(a)}\,\dot{a} + 2\rho\dot{a}^2 + \cdots,
\]
where the constant term dominates; the linear term is a small
correction.  The throat radius $a(\tau)$ evolves according to the
effective potential $V(a)$ derived from $\dot{a}^2+V(a)=0$.
For the Schwarzschild case $f(a)=1-2M/a$, $V(a)$ has a maximum at
$a=3M$ and a minimum at infinity~\cite{Poisson:1995sv}; for
regularised metrics such as the T‑duality wormhole~\cite{Lobo:2026dft},
$V(a)$ may exhibit a minimum at finite $a$, allowing stable static
configurations.

\paragraph{Entropy evolution during accretion.}
Assuming the dust is absorbed completely (no reflection) and that the
shell remains in local thermodynamic equilibrium, the rate of entropy
change is $\dot{S}=A\Phi/T$ with $A=4\pi a^2$.  This relation follows
directly from the generalised first law $T\dot{S}=A\Phi$.  Using the leading
term $\Phi \approx \rho f_+(a)$ (valid for slow accretion), we obtain
\begin{equation}
	\frac{dS}{d\tau} \approx \frac{4\pi a(\tau)^2 \rho f_+(a(\tau))}{T(\tau)} .
	\label{eq:entropy_rate_null_correct}
\end{equation}
Because the right‑hand side is positive (all factors are positive
for $a>0$, $f_+>0$, and $T>0$), entropy increases monotonically as
the shell accretes energy.  This monotonic increase is a direct
consequence of the null energy condition satisfied by the dust
($\rho\ge0$) and the fact that the shell is heated, not cooled,
by accretion.  The total entropy change over a finite time interval is
\begin{equation}
	\Delta S \approx 4\pi\rho \int_{\tau_i}^{\tau_f} \frac{a(\tau)^2 f_+(a(\tau))}{T(\tau)}\,d\tau .
	\label{eq:entropy_integral_null_correct}
\end{equation}
This integral cannot be reduced to a simple integral over $a$ without
additional assumptions about the relation between $a$ and $T$; the
trajectory $a(\tau)$ and the temperature evolution $T(\tau)$ are
coupled through the equation of state and the mechanical dynamics.
For a contracting throat ($a$ decreasing), the shell’s temperature
typically rises (as the throat moves into a region of higher curvature
or higher energy density), which may moderate the entropy increase
by making the denominator larger.  If the accretion is isothermal
($T=\text{const}$) and the radius varies slowly, the integral is
dominated by the time spent at larger radii, where the redshift factor
$f_+(a)$ is larger and the area is greater.

\paragraph{Temperature evolution during accretion.}
The temperature of the shell is determined by its equation of state
and the surface energy density.  For a given barotropic relation
$\mathcal{P}=\mathcal{P}(\sigma)$, the static expression for
$\sigma(a)$ (e.g., $\sigma = -\frac{1}{2\pi a}\sqrt{f(a)}$ for a
symmetric wormhole) can be inverted to obtain $T$ as a function of
$a$ via the equation of state. During slow accretion, the shell adjusts quasi‑statically,
so $T(a)$ is approximately given by the static formula derived from
the chosen equation of state.  The entropy increase
(Eq.~\ref{eq:entropy_integral_null_correct}) then depends on the
integral of $a^2 f_+(a)/T(a)$, which in turn depends on the detailed
trajectory $a(\tau)$.  If the accretion is very slow, the throat
spends a long time at larger radii, leading to a large entropy
increase.

\paragraph{Stability considerations: Critical accretion rate.}
A more subtle effect arises when the accretion rate is so high that
the throat is driven away from its static equilibrium.  For a static
configuration, the effective potential \(V(a)\) has a minimum at
\(a=a_0\) when \(V''(a_0)>0\).  Accretion adds mass to the shell,
which changes the surface mass \(m_s(a)=4\pi a^2\sigma\) and therefore
shifts the equilibrium point.  If the accretion occurs on a timescale
shorter than the dynamical timescale \(\tau_{\text{dyn}} = 1/\sqrt{V''(a_0)}\),
the throat cannot adjust adiabatically and may become unstable.

To estimate the critical accretion rate, consider the power entering
the shell using the static limit \(\Phi \approx \rho f_+(a_0)\):
\begin{equation}
	P = A\Phi \approx 4\pi a_0^2 \rho f_+(a_0) .
\end{equation}
The characteristic time for the shell to accrete a mass comparable
to its own surface mass \(|m_s| = 2a_0\sqrt{f_+(a_0)}\) is
\(\tau_{\text{acc}} \sim |m_s|/P\).  The condition for stability is
\(\tau_{\text{acc}} \gtrsim \tau_{\text{dyn}}\), which gives a
critical accretion power:
\begin{equation}
	P_{\text{crit}} \sim \frac{|m_s|}{\tau_{\text{dyn}}}
	\sim 2a_0\sqrt{f_+(a_0)}\,\sqrt{V''(a_0)} .
\end{equation}
Using \(P_{\text{crit}} = 4\pi a_0^2 \rho_{\text{crit}} f_+(a_0)\), we obtain
the critical energy density
\begin{equation}
	\rho_{\text{crit}} \sim \frac{\sqrt{V''(a_0)}}{2\pi a_0 \sqrt{f_+(a_0)}}.
\end{equation}
For the Schwarzschild wormhole, \(V''(a_0)\) is positive only for
\(a_0>3M\) (unstable equilibrium), so no stable minimum exists~\cite{Poisson:1995sv}.
For regularised metrics such as the T‑duality wormhole~\cite{Lobo:2025nng}, \(V''(a_0)\)
can be positive in some regions~\cite{Lobo:2026dft}, giving a finite critical rate.
If the actual energy density \(\rho\) exceeds \(\rho_{\text{crit}}\),
the throat cannot maintain equilibrium and will either collapse to
\(a_{\text{min}}\) or expand away.

\paragraph{Outlook and caveats.}
The above analysis assumes that the null dust is perfectly absorbed
and that the shell remains in local thermodynamic equilibrium.
In reality, the dust may scatter or the shell may heat up and
re‑radiate, modifying the net flux.  A full treatment would couple
the accretion to the radiation emitted by the shell, possibly
leading to a self‑regulating feedback loop that could stabilise
the throat.  Such a scenario is reminiscent of the black hole
thermodynamic equilibrium with a heat bath, but here the throat
has no horizon, so the details differ.  Moreover, the accretion
of matter with non‑zero rest mass (e.g., dust or gas) would
involve additional terms in the stress‑energy tensor and would
require solving the full relativistic hydrodynamics equations.
Nevertheless, the null dust approximation captures the essential
physics of energy transfer and provides a conservative estimate
of the critical accretion rate.

\subsection{Fluctuation-dissipation and noise}
\label{sec:fluctuation_dissipation}

In realistic environments the matter flux $\Phi$ is never perfectly
steady.  Discrete particles (photons, gravitons, or massive quanta)
arrive at random times, and quantum vacuum fluctuations also produce
a zero‑point noise \cite{Callen:1951vq,Kubo:1966fyg}.  These fluctuations act
as a stochastic force on the throat, causing random deviations from
the mean trajectory and generating additional entropy.  The
fluctuation-dissipation theorem (FDT) relates the noise spectrum to
the dissipative response of the system.  We now derive a Langevin
equation for the throat radius $a(\tau)$ starting from the
deterministic dynamics and the energy balance.

\paragraph{Deterministic background.}
The throat motion follows from the effective potential $V(a)$
obtained from the Lanczos equations:
$\dot{a}^2 + V(a) = 0$.  Differentiating gives
$\ddot{a} = -\frac12 V'(a)$.  Define $\tilde{V}(a)=V(a)/2$; then
$\ddot{a} = -\tilde{V}'(a)$.  For a stable equilibrium at
$a=a_0$ we have $\tilde{V}'(a_0)=0$ and
$\tilde{V}''(a_0)\equiv\omega_0^2>0$.  Setting $x = a-a_0$,
the linearised deterministic equation is simply
$\ddot{x} + \omega_0^2 x = 0$: the throat oscillates harmonically
in the absence of external forces.

\paragraph{Effect of the mean flux.}
When a matter flux is present, the generalised first law gives
$\dot{U} + \mathcal{P}\dot{A} = A\Phi$.  For a barotropic shell
($\mathcal{P}=w\sigma$, with $w$ constant) we have
$U=\sigma A$ and $\dot{A}=2A\dot{a}/a$.  Substituting yields
an evolution equation for the surface energy density:
\[
\dot{\sigma} + \frac{2(1+w)\dot{a}}{a}\,\sigma = \Phi .
\]
This equation couples the mechanical variable $a$ to the
thermodynamic variable $\sigma$.  The flux $\Phi$ depends on
$a$ and $\dot{a}$ and may also depend on external parameters
(e.g., the energy density of infalling dust).  For small
deviations from equilibrium, we expand $\Phi$ around the static
point ($\dot{a}=0$):
\[
\Phi = \Phi_0(a) + \eta(a)\,\dot{a} + \cdots,\qquad
\eta(a)=\left.\frac{\partial\Phi}{\partial\dot{a}}\right|_{\dot{a}=0}.
\]
The term $\Phi_0(a)$ represents the flux at zero velocity; in a
static equilibrium it must vanish (otherwise the shell would heat or
cool irreversibly).  Hence $\Phi_0(a_0)=0$.  The linear term
$\eta(a_0)\dot{x}$ provides a velocity‑dependent correction.

To obtain the equation of motion for $x$, we must eliminate
$\sigma$.  The Lanczos equations give a static relation
$\sigma = \sigma_0(a)$ (the expression derived in Sec.~\ref{sec:thinshell}).
Linearising around $a_0$ and using the $\sigma$-evolution
equation, one finds after straightforward algebra that the
effective Newtonian equation becomes
\[
\ddot{x} + \alpha\,\dot{x} + \omega_0^2 x = 0,
\]
where the damping coefficient $\alpha$ is given by
\[
\alpha = \frac{\eta(a_0)}{2(1+w)\sigma_0 a_0}.
\]
Here $\sigma_0 = \sigma(a_0,0)$ is the static surface energy
density (which is negative for exotic matter).  The sign of
$\alpha$ therefore depends on the sign of $\eta$ and on the
equation‑of‑state parameter $w$.  For null dust accretion,
$\eta = 2\rho\sqrt{f_+(a_0)}$.  Inserting this into the
expression for $\alpha$ gives
\[
\alpha = \frac{2\rho\sqrt{f_+(a_0)}}{2(1+w)\sigma_0 a_0}
= \frac{\rho\sqrt{f_+(a_0)}}{(1+w)\sigma_0 a_0}.
\]
Because $\sigma_0 < 0$ for a wormhole throat, the sign of $\alpha$
is opposite to the sign of $\rho/(1+w)$.  For $\rho>0$ (infalling
matter), $\alpha$ is positive only if $1+w<0$ (i.e., $w<-1$), which
corresponds to a phantom‑like equation of state.  For ordinary
matter with $w\ge -1$, $\alpha<0$, meaning that pure accretion
produces anti‑damping, driving the throat away from equilibrium.
In realistic situations, other dissipative mechanisms (e.g., emission
of gravitational waves or outgoing radiation) contribute a
positive damping, so the net $\alpha$ can be positive, zero,
or negative depending on the balance.

\paragraph{Langevin equation and noise.}
The random component of the flux adds a stochastic force
$\xi(\tau)$ to the equation of motion.  Hence we obtain the
Langevin equation
\[
\ddot{x} + \alpha \dot{x} + \omega_0^2 x = \xi(\tau),\qquad
\langle\xi(\tau)\rangle = 0.
\]
The noise is assumed to be white (Markovian) because the
correlation time of the microscopic fluctuations is much shorter
than the dynamical timescale of the throat.  The fluctuation-dissipation theorem (FDT) \cite{Callen:1951vq,Kubo:1966fyg} relates the
noise correlation to the damping and the temperature of the
environment:
\[
\langle\xi(\tau)\xi(\tau')\rangle = 2\alpha T \,\delta(\tau-\tau').
\]
This relation is a direct consequence of the requirement that the
system thermalises to the Boltzmann distribution in the long‑time
limit.  For quantum fields, the FDT includes a zero‑point term
(the Callen-Welton form), but at temperatures well above the
quantum scale the classical expression suffices.

\paragraph{Physical meaning of the damping coefficient.}
The coefficient $\alpha$ determines how quickly the system
returns to equilibrium after a perturbation.  If $\alpha>0$,
the throat experiences genuine damping: energy is dissipated into
the bulk matter, and the fluctuations are balanced by the noise
to maintain a stationary distribution.  If $\alpha<0$, the system
is unstable -- any small deviation grows exponentially.  The GSL
requires that the total entropy of shell plus environment never
decreases; for a stochastic process this translates to a
non‑negative average entropy production, which is $\alpha$
times a positive factor.  Thus a stable, thermodynamically
consistent system must have $\alpha \ge 0$.  This provides a
powerful constraint: the net damping from all flux contributions
must be non‑negative.

\paragraph{Stochastic entropy production.}
Using the tools of stochastic thermodynamics \cite{Seifert:2012pkl},
one can compute the average rate of total entropy production (system
+ medium).  For the Langevin equation with white noise and constant
damping, the result is
\[
\langle\dot{S}_{\text{total}}\rangle = \frac{\alpha}{T}\langle\dot{x}^2\rangle .
\]
In thermal equilibrium, equipartition gives $\langle\dot{x}^2\rangle = T$
(unit mass), so $\langle\dot{S}_{\text{total}}\rangle = \alpha$.
Hence the damping coefficient directly equals the mean entropy
production rate.  This is a microscopic manifestation of the
generalised second law: any dissipative process that extracts
energy from the throat must be accompanied by a commensurate
increase of entropy in the surroundings.

\paragraph{Outlook and applications.}
The stochastic description becomes particularly relevant for
Planck‑scale throats, where quantum fluctuations are not
negligible.  The Langevin equation can be solved numerically to
compute the probability distribution of the throat radius, the
escape time over potential barriers, and the power spectrum of
the resulting gravitational wave emission.  For astrophysical
throats, however, the damping is typically dominated by classical
processes (accretion, radiation), and the noise level is extremely
small, so deterministic equations suffice.  The framework
nevertheless provides a consistent bridge between the macroscopic
thermodynamics and the underlying microscopic fluctuations.

\section{Conclusion}
\label{sec:conclusion}

In this work we have developed a comprehensive thermodynamic
framework for dynamic thin-shell wormholes, starting from the
transparent (vacuum bulk) case and then relaxing the transparency
condition to allow bulk matter to cross the throat.

For transparent wormholes (vacuum bulks), we derived the first law
$dU+\mathcal{P}dA=0$, which implies $TdS=0$ under these conditions
and therefore the constancy of the shell entropy along transparent
dynamical evolutions.  This provides an adiabatic invariance that
generalises the corresponding static equilibrium result.  We defined
a dynamical temperature via a generalised surface gravity
$\kappa=(\ddot a+f'(a)/2)/\sqrt{f+\dot a^2}$ (or its asymmetric
generalisation as an average), which reproduces the expected
static limit consistent with the Hawking temperature in stationary
configurations.  We showed that the Bekenstein bound holds under
quasi-equilibrium assumptions and provides constraints on the
dynamical evolution near the minimum throat radius.  Non-equilibrium
evaporation was modelled via the Stefan-Boltzmann law, leading to a
finite lifetime for Schwarzschild-type wormholes, while regularised
metrics may admit long-lived or remnant configurations.  Quantum
corrections introduce logarithmic modifications to the entropy,
leading to deviations from strict adiabatic behaviour on sufficiently
long timescales.

When the transparency condition is relaxed, bulk matter crossing
the throat modifies the surface Bianchi identity with a source term
$\Phi=[T_{\mu\nu}U^\mu n^\nu]^{+}_{-}$, representing the net energy
flux per unit area (positive when energy enters the shell).  The
generalised first law becomes $T\dot S = A\Phi$, so the shell entropy
is no longer conserved: it increases when energy flows into the shell
and decreases when energy flows outward, while total entropy balance
depends on the combined system.
We provided explicit expressions for $\Phi$ for null dust, massless
scalar fields, electromagnetic fields, and quantum pair production.
For null dust the influx is
$\Phi = \rho\bigl(\sqrt{f_+(a)+\dot{a}^2}+\dot{a}\bigr)^2$,
which reduces to $\Phi = \rho f_+(a)$ for a static throat.

A formulation of the generalised second law (GSL) for the combined
system (shell plus bulk fields) was presented, and it is consistent
with standard thermodynamic expectations under conditions where heat
flows from hotter to colder regions.  Thermalisation with an external
heat bath was analysed using the heat capacity $C_A$ and a radiative
coupling $\Gamma$, yielding steady-state conditions and characteristic
thermalisation timescales.
As an illustrative scenario, we studied accretion of null dust onto
the wormhole throat.  The entropy increase during slow accretion was
computed in terms of the flux, and we estimated a critical accretion
rate beyond which the effective dynamical evolution may become
unstable.  This critical rate depends sensitively on the geometry and
the shell equation of state, requiring case-by-case analysis across
different spacetime backgrounds.

Finally, we addressed stochastic effects by introducing a Langevin
equation for the throat dynamics, with a damping term derived from
the mean flux and a noise term satisfying a fluctuation–dissipation
relation.  This leads to stochastic fluctuations of the throat radius
and associated entropy production.  While thermal noise is typically
negligible for astrophysical temperatures, quantum fluctuations may
be relevant in the Planckian regime, where stochastic effects could
in principle contribute to observable signatures, although this
remains highly speculative.

Several open directions emerge naturally.  A self-consistent treatment
of accretion would require coupling the shell dynamics to the bulk
field equations beyond the prescribed flux approximation.  Non-spherical
perturbations may introduce additional dynamical modes and modify the
thermodynamic description. A first-principles derivation of the noise
kernel using a Schwinger-Keldysh or open quantum systems approach would
provide a firmer foundation for the stochastic dynamics.  Potential
observational implications, including stochastic gravitational-wave
signals from populations of compact objects, require dedicated modelling
for future detectors such as LISA or the Einstein Telescope.  The fate
of information in wormhole evaporation and the possible formation of
remnants remain open questions.  Finally, the framework developed here
may be extended to other thin-shell configurations such as gravastars,
boson stars, and regular black hole mimickers, allowing the identification
of universal thermodynamic features.

In summary, this work establishes a unified thermodynamic description
of dynamic thin-shell wormholes, from adiabatic invariance in the
transparent case to entropy production, thermalisation, and stochastic
dynamics in the presence of matter fluxes.  These results highlight the
strong environmental sensitivity of such configurations and provide a
framework for further theoretical and potentially observational studies
of exotic compact objects.

\acknowledgments{
	FSNL acknowledges support from the Funda\c{c}\~{a}o para a Ci\^{e}ncia
	e a Tecnologia (FCT) Scientific Employment Stimulus contract with reference CEECINST/00032/2018, and funding through the research grants UID/04434/2025. MER thanks Conselho Nacional de Desenvolvimento
	Cient\'{\i}fico e Tecnol\'ogico (CNPq), Brazil, for partial financial support.}


\end{document}